\documentclass[12pt]{article}
\def\ffrac#1#2{\textstyle{#1\over#2}\displaystyle}

\usepackage{psfig,epsfig,amsfonts,latexsym}

\def\Ladder{
\unitlength=0.50mm
\begin{picture}(160.,50.)
\put(0,0){\line(1,0){150.}}
\put(0,30){\line(1,0){150.}}
\put(155.,0.){\makebox(0.,0.){$S$}}
\put(155.,30.){\makebox(0.,0.){$T$}}
\put(-7.,0.){\makebox(0.,0.){$g_s$}}
\put(-7.,30.){\makebox(0.,0.){$g_t$}}
\put(30,0){\line(0,1){30.}}
\put(60,0){\line(0,1){30.}}
\put(90,0){\line(0,1){30.}}
\put(120,0){\line(0,1){30.}}
\put(30.,0.){\makebox(0.,0.){$\bullet$}}
\put(30.,30.){\makebox(0.,0.){$\bullet$}}
\put(60.,0.){\makebox(0.,0.){$\bullet$}}
\put(60.,30.){\makebox(0.,0.){$\bullet$}}
\put(90.,0.){\makebox(0.,0.){$\bullet$}}
\put(90.,30.){\makebox(0.,0.){$\bullet$}}
\put(120.,0.){\makebox(0.,0.){$\bullet$}}
\put(120.,30.){\makebox(0.,0.){$\bullet$}}
\put(63.,14.){\makebox(0.,0.){$J_{\perp}$}}
\put(74.,25.){\makebox(0.,0.){$J_{\parallel}$}}
\put(74.,2.){\makebox(0.,0.){$J_{\parallel}$}}
\put(84.,14.){\makebox(0.,0.){$J_{\perp}$}}
\end{picture}}

\def\LaddF{
\unitlength=0.50mm
\begin{picture}(160.,50.)  
\put(0,0){\line(1,0){150.}}
\put(0,30){\line(1,0){150.}}
\put(155.,0.){\makebox(0.,0.){$S$}}
\put(155.,30.){\makebox(0.,0.){$T$}}
\put(30,5){\line(0,1){20.}}  
\put(60,5){\line(0,1){20.}}
\put(90,5){\line(0,1){20.}}  
\put(120,5){\line(0,1){20.}}
\put(30.,30.){\oval(10.,10.)[l]}
\put(30.,30.){\oval(10.,10.)[r]}
\put(30.,27.5){\vector(0,1){5}}
\put(30.,0.){\oval(10.,10.)[l]}
\put(30.,0.){\oval(10.,10.)[r]}
\put(27.5,-2.5){\vector(0,1){5}}
\put(32.5,2.5){\vector(0,-1){5}}
\put(60.,30.){\oval(10.,10.)[l]}
\put(60.,30.){\oval(10.,10.)[r]}
\put(60.,32.5){\vector(0,-1){5}}
\put(60.,0.){\oval(10.,10.)[l]}
\put(60.,0.){\oval(10.,10.)[r]}
\put(57.5,-2.5){\vector(0,1){5}}
\put(62.5,-2.5){\vector(0,1){5}}
\put(90.,30.){\oval(10.,10.)[l]}
\put(90.,30.){\oval(10.,10.)[r]}
\put(90.,27.5){\vector(0,1){5}}
\put(90.,0.){\oval(10.,10.)[l]}
\put(90.,0.){\oval(10.,10.)[r]}
\put(87.5,-2.5){\vector(0,1){5}}
\put(92.5,2.5){\vector(0,-1){5}}
\put(120.,30.){\oval(10.,10.)[l]}
\put(120.,30.){\oval(10.,10.)[r]}
\put(120.,32.5){\vector(0,-1){5}}
\put(120.,0.){\oval(10.,10.)[l]}
\put(120.,0.){\oval(10.,10.)[r]}
\put(117.5,-2.5){\vector(0,1){5}}
\put(122.5,-2.5){\vector(0,1){5}}
\end{picture}}

\def\Ladderg2{
\unitlength=0.50mm
\begin{picture}(160.,50.)  
\put(0,0){\line(1,0){150.}}
\put(0,30){\line(1,0){150.}}
\put(155.,0.){\makebox(0.,0.){$S$}}
\put(155.,30.){\makebox(0.,0.){$T$}}
\put(30,5){\line(0,1){20.}}  
\put(60,5){\line(0,1){20.}}
\put(90,5){\line(0,1){20.}}  
\put(120,5){\line(0,1){20.}}
\put(30.,30.){\oval(10.,10.)[l]}
\put(30.,30.){\oval(10.,10.)[r]}
\put(30.,27.5){\vector(0,1){5}}
\put(30.,0.){\oval(10.,10.)[l]}
\put(30.,0.){\oval(10.,10.)[r]}
\put(27.5,-2.5){\vector(0,1){5}}
\put(32.5,-2.5){\vector(0,1){5}}
\put(60.,30.){\oval(10.,10.)[l]}
\put(60.,30.){\oval(10.,10.)[r]}
\put(60.,27.5){\vector(0,1){5}}
\put(60.,0.){\oval(10.,10.)[l]}
\put(60.,0.){\oval(10.,10.)[r]}
\put(57.5,-2.5){\vector(0,1){5}}
\put(62.5,-2.5){\vector(0,1){5}}
\put(90.,30.){\oval(10.,10.)[l]}
\put(90.,30.){\oval(10.,10.)[r]}
\put(90.,27.5){\vector(0,1){5}}
\put(90.,0.){\oval(10.,10.)[l]}
\put(90.,0.){\oval(10.,10.)[r]}
\put(87.5,-2.5){\vector(0,1){5}}
\put(92.5,-2.5){\vector(0,1){5}}
\put(120.,30.){\oval(10.,10.)[l]}
\put(120.,30.){\oval(10.,10.)[r]}
\put(120.,27.5){\vector(0,1){5}}
\put(120.,0.){\oval(10.,10.)[l]}
\put(120.,0.){\oval(10.,10.)[r]}
\put(117.5,-2.5){\vector(0,1){5}}
\put(122.5,-2.5){\vector(0,1){5}}
\end{picture}}

\begin{document}

\begin{center}
{\bf Quantum phase diagram of an exactly solved mixed spin ladder}\\
\vskip 3mm
{M.T. Batchelor$^{\dag}$, X.-W. Guan$^{\dag}$, N. Oelkers$^{\dag}$ 
and  Z.-J. Ying$^{\ddag}$}\\
\vskip 3mm
${}^{\dag}$ Department of Theoretical Physics,\\ 
Research School of Physical Sciences and Engineering,\\
and\\
Centre for Mathematics and its Applications,\\
Mathematical Sciences Institute,\\
Australian National University, Canberra ACT 0200,  Australia\\
\vskip 3mm
$^{\ddag}$Instituto de F\'{\i}sica da UFRGS,
                     Av.\ Bento Gon\c{c}alves, 9500,\\
                     Porto Alegre, 91501-970, Brasil\\
and\\
Hangzhou Teachers College, Hangzhou 310012, China

\end{center}

\begin{abstract}
We investigate the quantum phase diagram of the exactly solved mixed spin-$(\frac{1}{2},1)$ ladder
via the thermodynamic Bethe ansatz (TBA). In the absence of a magnetic field the model exhibits three
quantum phases associated with  $su(2)$, $su(4)$ and $su(6)$ symmetries.  In the presence of a strong
magnetic field, there is a third and full saturation magnetization plateaux within the strong antiferromagnetic
rung coupling regime. Gapless and gapped phases appear in turn as the magnetic field increases. For weak
rung coupling, the fractional magnetization plateau vanishs and exhibits new quantum phase transitions.
However, in the ferromagnetic coupling regime, the system does not have a third saturation magnetization plateau.
The critical behaviour in the vicinity of the critical points is also derived systematically using the TBA.
\end{abstract}
\newpage
\section{Introduction} \label{sec1}

The field of exactly solved models in statistical mechanics has many significant highlights.
These include Elliott Lieb's pioneering work on the six-vertex model and his calculation of the
residual entropy of square ice \cite{ice}.
Over the ensuing years the six-vertex model and the related Heisenberg spin chain have been
generalised in all manner of directions.
Most recently attention has turned to the physics of quantum spin ladders,
for which a number of exactly solved models have been proposed.
The underlying model is the spin-$\frac{1}{2}$ Heisenberg ladder, which has been
studied extensively \cite{exp1}.
This model consists of two Heisenberg chains coupled together with Heisenberg rung
interactions forming a ladder-like structure.
A number of ladder compounds have been synthesized, such as  SrCu$_2$O$_3$ \cite{exp2},
Cu$_{2}$(C$_5$H$_{12}$N$_2$)$_2$Cl$_4$ \cite{exp3},
(C$_5$H$_{12}$N)$_2$CuBr$_{4}$ \cite{exp4}, (5IAP)$_2$CuBr$_4\cdot
2$H$_2$O \cite{exp5}, KCuCl$_3$ and TlCuCl$_3$ \cite{exp6}.
The experimental results reveal an interesting mix of low-temperature physics, including
spin excitation gaps and magnetization plateaux.

The theoretical investigation of the ladder compounds has also been centred on a number of variants 
of the standard Heisenberg ladder \cite{Hlad1}, including the
addition of multi-body interactions \cite{ladd1}, alternation and frustration \cite{alt,fra}.
The ladder models have been studied by a variety of methods, for example, 
numerical \cite{Hlad1,numer1}, perturbation theory \cite{FT1, FT2} and
the quantum transfer matrix algorithm \cite{troyer}.
Unfortunately the Heisenberg ladder is not exactly solvable in the sense of the 
six-vertex model or related $su(2)$ Heisenberg chain.
However, a number of variants have been solved exactly by means of the Bethe ansatz; 
see, for example, Refs \cite{ladd4,ladd2,ladd3,mlad2,ladd5,ladd6}.
For arguably the simplest model, based on $su(4)$, the critical behaviour derived from the
thermodynamic Bethe ansatz (TBA) is seen to be consistent with the existing experimental, 
numerical and perturbative results for the strong coupling ladder compounds \cite{BGFZ}.
This includes the spin excitation gap and the critical fields $H_{c1}$ and $H_{c2}$, which
are in excellent agreement with the experimental values for the known strong coupling
ladder compounds (5IAP)$_2$CuBr$_4\cdot 2$H$_2$O, Cu$_{2}$(C$_5$H$_{12}$N$_2$)$_2$Cl$_4$ 
and (C$_5$H$_{12}$N)$_2$CuBr$_{4}$.

On the other hand, the special interest in fractional magnetization plateaux \cite{FPL} has
inspired work on  mixed spin chains \cite{mix1,mix2,mix3,mix4}, mixed spin ladders
\cite{mlad1} and various experimental compounds \cite{MC1,MC2,mixladd}.
In particular, the magnetic behaviour of a mixed spin-$(\frac{1}{2},1)$ Heisenberg ladder 
has been investigated by means of the density-matrix renormalization group
technique \cite{mlad1}.
It was concluded that for certain strong rung coupling magnetization plateaux exist
at $M^z=0.5$ and at $M^z=1$, but with no plateaux for negative (ferromagnetic) rung coupling.
In fact the mixed spin ladder exhibits a richer phase diagram than the spin-$\frac{1}{2}$ ladder.
It appears that  a comprehensive study of the mixed spin ladder,  e.g., 
the prediction of the critical fields for different rung coupling and examination
of the critical behavior, has not been undertaken.
It also remains to investigate the effect of non-equal Land\'{e} $g$-factors on the ladder legs,
particularly given that $g$-factor anisotropy appears to affect the critical fields in the 
spin-orbital model \cite{SO1,SO2}.

In this paper, we investigate the quantum phase diagram of the exactly solved mixed
spin-$(\frac{1}{2},1)$ ladder \cite{mlad2}, with different $g$-factors, via the TBA. We find
that in the absence of a magnetic field the model exhibits three
quantum phases associated with $su(2)$, $su(4)$ and $su(6)$ symmetries.
In the presence of a strong magnetic field $h$, two
magnetization plateaux appear in the strong antiferromagnetic rung coupling regime. 
The fractional magnetization  plateau $M^z=\frac{1}{2}g_s\mu
_B+\frac{1}{6}(g_s-g_t) \mu_B$ corresponding to the fully-polarized doublet
rung state, opens at the critical field $H_{c1} $ and vanishes at the
critical field $H_{c2} $.  The second plateau
$M^z=(\frac{1}{2}g_t+g_s)\mu_B$, corresponding to a  fully-polarized
quadruplet rung state, opens only at a very strong magnetic field $H_{c3}$.
For weak antiferromagnetic rung coupling, the fractional plateau is closed
such that three different kinds of quantum phase transition occur.
For ferromagnetic rung coupling only the
full saturation magnetization plateau exists. 
The critical behavior for the different quantum phase transitions is
systematically derived by using the TBA.

The paper is organized as follows.
 In section \ref{sec2}, we discuss the exactly solved
mixed spin-$(\frac{1}{2},1)$ ladder model with different
Land\'{e} factors. The exact solution is given via an appropriate choice
of rung basis.  Section \ref{sec3} is devoted to the investigation of
the quantum phase diagram. 
In Section \ref{sec4} we investigate the magnetization plateaux in
the presence of a strong magnetic field. The critical fields
characterizing the different quantum phase transitions are given
explicitly. A summary of our main results and conclusions are given in Section \ref{sec5}.

\section{The exactly solved mixed spin-$(\frac{1}{2},1)$ ladder model}
\label{sec2}

The Hamiltonian of the exactly solved spin-$(\frac{1}{2},1)$ ladder model,  based on the $su(6)$ 
symmetry, reads \cite{mlad2}
\begin{equation}
H=J_{\parallel}H_{{\rm
leg}}+J_{\perp}\sum_{j=1}^{L}\vec{T}_j \cdot \vec{S}_j -g_t\mu_{{\rm
B}}h\sum_{j=1}^{L}T^z_j-g_s\mu_{{\rm
B}}h\sum_{j=1}^{L}S^z_j,\label{Ham}
\end{equation}
\begin{equation}
H_{{\rm
leg}}=\sum_{j=1}^{L}\left(\frac{1}{2}+2\, \vec{T}_j \cdot \vec{T}_{j+1}\right)
\left(-1+\vec{S}_j \cdot \vec{S}_{j+1}+(\vec{S}_j \cdot \vec{S}_{j+1})^2\right).
\label{intra}
\end{equation}
Here $\vec{T }_j$ and $\vec{S}_j$ are
the standard spin-$\frac{1}{2}$ and spin-$1$ operators acting on 
site $j$ of the upper and lower legs, respectively (see Fig.1),
$J_{\parallel}$ and $J_{\perp}$ are the intrachain and interchain coupling constants, $h$ is
the magnetic field, $\mu_{{\rm B}}$ is the Bohr magneton and $g_t$ and
$g_s$ are the Land\'{e} factors along each leg.
Periodic boundary conditions are imposed with $L$ the number of rungs.

\begin{figure}[t]
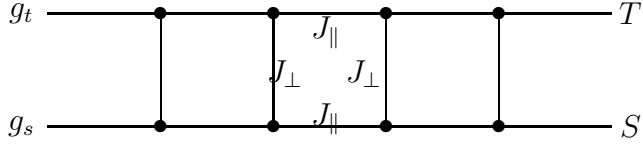

\begin{center}
\Ladder \\
\end{center}
\caption{The mixed  spin ladder: $J_{\parallel}$ and $J_{\perp}$ are the intrachain and 
interchain couplings; $g_t$ and $g_s$ are the Land\'{e} factors along each leg.}
\end{figure}

The rung term in (\ref{Ham}) breaks the $su(6)$ symmetry of $H_{{\rm leg}}$ into
$su(4)\oplus u(2)$ symmetry. 
This symmetry is in turn broken under the magnetic field.
The physical properties and the critical behaviour of the model are determined
by the competition between the rung and leg coupling constants and the magnetic field $h$.
Essentially,  $H_{{\rm leg}}$ is the permutation operator
corresponding to the $su(6)$ algebra symmetry. 
If we change the
canonical basis $e_i^{\alpha \beta}\otimes e_j^{\gamma \delta}$ of
$V_1\otimes V_2$ into rung quadruplet and doublet states (Clebsch-Gordon
decomposition), the six-dimensional space splits into the
direct sum of quadruplets and doublets with the basis
\begin{eqnarray}
|1\rangle
&=&\ffrac{\sqrt{2}}{\sqrt{3}}\left(|1,-\ffrac{1}{2}\rangle-\ffrac{1}{\sqrt{2}}|0,\ffrac{1}{2}
\rangle\right),\quad
|2\rangle
=\ffrac{\sqrt{2}}{\sqrt{3}}\left(|-1,\ffrac{1}{2}\rangle-\ffrac{1}{\sqrt{2}}|0,-\ffrac{1}{2}
\rangle\right),\nonumber\\
|3\rangle &=&|1,\ffrac{1}{2}\rangle,\quad |4\rangle
=\ffrac{1}{\sqrt{3}}\left(|1,-\ffrac{1}{2}\rangle+\sqrt{2}|0,\ffrac{1}{2}\rangle\right),\nonumber\\
|5\rangle
&=&\ffrac{1}{\sqrt{3}}\left(|-1,\ffrac{1}{2}\rangle+\sqrt{2}|0,-\ffrac{1}{2}\rangle\right),\,\,|6\rangle
=|-1,-\ffrac{1}{2}\rangle, \label{eigB}
\end{eqnarray}
where the states
$|1\rangle,\, |2\rangle$ form the doublet and the remaining
states form the quadruplet. The projectors onto the doublet and
quadruplet subspace are given by
\begin{equation} P_{{\rm
d}}=-\ffrac{2}{3}(\vec{T} \cdot \vec{S}-\ffrac{1}{2}),\,\, P_{{\rm
q}}=\ffrac{2}{3}(\vec{T} \cdot \vec{S}+1).
\end{equation}
It follows that the rung
interaction term can be accommodated into an $su(6)$ invariant Heisenberg
chain by embedding the doublet rung states through an appropriate chemical
potential term. The leg and rung part of the Hamiltonian (\ref{Ham})
can be derived from the relation 
\begin{equation}
H=J_{\parallel}\frac{d}{dv}\ln \tau (v)|_{v=0}+E_{\perp+h}+ {\rm
const}\label{H-T}
\end{equation}
associated with the quantum transfer matrix $\tau (v)={\rm tr}_0T(v)$.
The energy $E_{\perp+h}$ arising from the rung interaction and magnetic field terms
is given further below. 
Here $T(u)$ denotes the monodromy matrix given by
\begin{equation}
T(v )= R_{0,L}(v)R_{0,L-1}(v )\ldots R_{0,2}(v)R_{0,1}(v)
\end{equation}
associated with the $su(6)$ quantum $R$-matrix.  

Now consider the effect of the magnetic field.
Although the magnetic field preserves the integrability of the leg part of the
Hamiltonian, the different $g$-factors on each leg break the 
doublet/quadruplet basis (\ref{eigB}) for the Hamiltonian (\ref{Ham}).
Fortunately, we can still find another basis,
\begin{eqnarray}
\psi^{(\pm)}_{\frac{1}{2}}&=&\frac{1}{\sqrt{1+(y^{(\pm)}_{\frac{1}{2}})^2}}
\left(|1,-\ffrac{1}{2}\rangle
+y^{(\pm)}_{\frac{1}{2}}|0,\ffrac{1}{2}\rangle\right),\nonumber\\
\psi^{(\pm)}_{-\frac{1}{2}}&=&\frac{1}{\sqrt{1+(y^{(\pm)}_{-\frac{1}{2}})^2}}
\left(|-1,\ffrac{1}{2}\rangle
+y^{(\pm)}_{-\frac{1}{2}}|0,-\ffrac{1}{2}\rangle\right), \\
\psi_{\frac{3}{2}}&=&|1,\ffrac{1}{2}\rangle,
\,\,\psi_{-\frac{3}{2}}=|-1,-\ffrac{1}{2}\rangle, \nonumber
\end{eqnarray}
to diagonalize the rung and magnetization terms simultaneously.
The quantities $y^{(\pm)}_{a}$ are given by
\begin{eqnarray}
 y^{(\pm)}_{a}=a\sqrt{2}[(g_s-g_t)h^{'}+a] \pm
\sqrt{1+\ffrac{1}{2}(g_sh^{'}-g_th^{'}+a)^2}, 
\end{eqnarray}
where $a=\pm \frac{1}{2}$ and $h^{'}=\mu_B h/J_{\perp}$.
We notice that if $g_s=g_t$, the basis states $\psi^{(-)}_{\frac{1}{2}}$
and $\psi^{(-)}_{-\frac{1}{2}}$ reduce to the doublet, with
the other states reducing to the quadruplet.
With regard to the total spin of the multiplets we can still call the  states 
$(\psi_{\frac{3}{2}},\psi_{\frac{1}{2}}^{(+)},\psi_{-\frac{1}{2}}^{(+)},\psi_{-\frac{3}{2}})$
quadruplets and $(\psi^{(-)}_{\frac{1}{2}},\psi^{(-)}_{-\frac{1}{2}})$ doublets.

It is well established that the Hamiltonian (\ref{Ham}) can be diagonalized via the 
algebraic Bethe ansatz. 
In this procedure it is important to note that the leg content of the Hamiltonian, $H_{{\rm leg}}$, 
is not altered under the change of basis  order between the quadruplet and the doublet states,
however the rung and magnetic field terms are altered by these changes.
We note also that for the ladder Hamiltonian (\ref{Ham}) the doublet rung state is energetically 
favoured for $J_{\perp}>0$, whereas the quadruplet rung state is favoured for $J_{\perp}<0$. 
This is the reason for choosing the doublet  state as reference state for $J_{\perp}>0$, while a
 quadruplet state is choosen as reference state for $J_{\perp}<0$. 
As the magnetic field is turned on, the energy levels of each multiplet component split. 
The basis order is therefore chosen in accordance with their energy levels.

The resulting Bethe ansatz equations are well known \cite{BA} and
consist of a set of five coupled equations depending on five flavours, $v^{(k)},\,k=1,\ldots,5$. 
The Bethe ansatz equations 
\begin{equation}
\prod_{i=1}^{M_{k-1}}\frac{v_j^{(k)}-v_i^{(k-1)}+\frac{\mathrm{i}}{2}}
{v_j^{(k)}-v_i^{(k-1)}-\frac{\mathrm{i}}{2}}=\prod^{M_k}_{\stackrel{\scriptstyle l=1}{l\neq j}}
\frac{v_j^{(k)}-v_l^{(k)}+\mathrm{i}}{v_j^{(k)}-v_l^{(k)}-\mathrm{i}}
\prod_{l=1}^{M_{k+1}}\frac{v_j^{(k)}-v_l^{(k+1)}-\frac{\mathrm{i}}{2}}
{v_j^{(k)}-v_l^{(k+1)}+\frac{\mathrm{i}}{2}}\label{BE}
\end{equation}
can be derived from the nested algebraic Bethe ansatz.
In the above, $k=1,\ldots,5$ and $ j=1,\ldots,M_k$ and the conventions
$v_j^{(0)}=v_j^{(6)}=0,\, M_6=0$ apply. 
In real ladder compounds the difference between
the $g$-factors along each leg is small (this is not always true for spin-orbital
models \cite{SO1}). 
Thus from now on we treat the term  $g_s-g_t$ as a small quantity.  
After some algebra, the eigenspectrum is obtained from relation (\ref{H-T}) as
\begin{equation}
E=J_{\parallel}L-J_{\parallel}\sum^{M_1}_{i=1}\frac{1}{v_i^2+\frac{1}{4}}+E_{\perp+h},
\end{equation}
where the energy contribution from the rung interaction and
the magnetic field terms is given  by
\begin{eqnarray}
E_{\perp+h} &=&\left[-\ffrac14 J_{\perp} -\ffrac{1}{2} g_s\mu_Bh-\ffrac{1}{\sqrt 2} J_{\perp} 
\sqrt{1+\ffrac{1}{2}(g_sh^{'}-g_th^{'}+\ffrac{1}{2})^2}\right]N^{(-)}_{\frac{1}{2}}\nonumber\\
& &
+\left[-\ffrac14 J_{\perp} +\ffrac{1}{2}g_s\mu_Bh-\ffrac{1}{\sqrt 2} J_{\perp} 
\sqrt{1+\ffrac{1}{2}(g_sh^{'}-g_th^{'}-\ffrac{1}{2})^2}\right]N^{(-)}_{-\frac{1}{2}}\nonumber\\
& &+\left[\ffrac12 J_{\perp} -(\ffrac{1}{2}g_t+g_s)\mu_Bh\right]N_{\frac{3}{2}}\nonumber\\
& &\left[-\ffrac14 J_{\perp} -\ffrac{1}{2}g_s\mu_Bh+\ffrac{1}{\sqrt 2} J_{\perp} 
\sqrt{1+\ffrac{1}{2}(g_sh^{'}-g_th^{'}+\ffrac{1}{2})^2}\right]N^{(+)}_{\frac{1}{2}}\nonumber\\
& &+\left[-\ffrac14 J_{\perp}+\ffrac{1}{2}g_s\mu_Bh+\ffrac{1}{\sqrt 2} J_{\perp} 
\sqrt{1+\ffrac{1}{2}(g_sh^{'}-g_th^{'}-\ffrac{1}{2})^2}\right]N^{(+)}_{-\frac{1}{2}}\nonumber\\
& &+[\ffrac12 J_{\perp}+(\ffrac{1}{2}g_t+g_s)\mu_Bh]N_{-\frac{3}{2}}\nonumber\\
&=&[-\ffrac{3}{2}J_{\perp}-\ffrac{1}{2}g_s\mu_Bh-\ffrac{1}{6}(g_s-g_t)
\mu_Bh]N^{(-)}_{\frac{1}{2}}\nonumber\\
& &+[-\ffrac{3}{2}J_{\perp}+\ffrac{1}{2}g_s\mu_Bh+\ffrac{1}{6}(g_s-g_t)
\mu_Bh]N^{(-)}_{-\frac{1}{2}}\nonumber\\
& &-(\ffrac{1}{2}g_t+g_s)\mu_BhN_{\frac{3}{2}}
-[\ffrac{1}{2}g_s\mu_Bh-\ffrac{1}{6}(g_s-g_t)
\mu_Bh]N^{(+)}_{\frac{1}{2}}\nonumber\\
& &+[\ffrac{1}{2}g_s\mu_Bh-\ffrac{1}{6}(g_s-g_t)
\mu_Bh]N^{(+)}_{-\frac{1}{2}}\nonumber\\
& &+(\ffrac{1}{2}g_t+g_s)\mu_BhN_{-\frac{3}{2}}+ {\rm const.}\label{energy}
\end{eqnarray}
Here the $N$'s are the numbers of the corresponding states.
In the thermodynamic limit, the Bethe ansatz equations (\ref{BE}) allow
the string solution \cite{TBA,Lee}
\begin{eqnarray}
{v^{(k)}}^n_{\alpha_k j}&=&{v^{(k)}}^n_{\alpha_1}+\ffrac12 \mathrm{i} (n+1-2j),
\end{eqnarray}
where $j=1,...,n$, $\alpha_a =1,...,N_n^{(k)}$ and ${v^{(k)}}^n_{\alpha_k},\,k=1,...,5$, are  the
positions of the center of the strings of flavour $k$. The number
of $n$-strings, $N_n^{(a)}$, satisfies the relation $M^{(k)}=\sum_nnN_n^{(k)}$.
On taking the thermodynamic limit, the Bethe ansatz equations become
\begin{eqnarray}
\rho^{(1)h}_n&=&a_n-\sum_{m}A_{nm}*\rho^{(1)}_m+\sum_{m}a_{nm}*\rho^{(2)}_m,\label{Tbethe1}\\
\rho^{(k)h}_n&=&-\sum_{m}A_{nm}*\rho^{(k)}_m+\sum_{m}a_{nm}*(\rho^{(k-1)}_m+\rho^{(k+1)}_m),
\label{Tbethe2}
\end{eqnarray}
where $k=2,\ldots,5$, and the symbol $*$ denotes convolution. 
$\rho^{(k)}_n(v)$ and $\rho^{(k)h}_n(v)$ with $ k=1,\ldots,5$ 
are the densities of roots and holes for the five flavours.
We have adopted the standard notations
\begin{eqnarray}
A_{nm}(\lambda)&=& \delta(\lambda)\delta_{nm}+(1-\delta_{nm})a_{|n-m|}(\lambda)
+a_{n+m}(\lambda) \\
& &+2\sum^{{\rm Min}(n,m)-1}_{l=1}a_{|n-m|+2l}(\lambda),\nonumber\\
a_{nm}(\lambda)& =& \sum^{{\rm Min}(n,m)}_{l=1}a_{n+m+1-2l}(\lambda), 
\end{eqnarray}
with $a_n(\lambda)=\frac{1}{2\pi}\frac{n}{n^2/4+\lambda ^2}$.

In order to find the equilibrium state of the system at a fixed
temperature $T$ and external magnetic field $h \geq 0$, we
minimize  the free energy $F=E-TS-hM^z$ with respect to the densities and then obtain the TBA 
equations in the form
\begin{eqnarray}
\epsilon^{(k)}_1&=&g_1^{(k)}+Ta_2*\ln(1+e^{-\frac{\epsilon^{(k)}_1}{T}})+
T(a_0+a_2)\sum_{m=1}^{\infty}a_m*\ln(1+e^{-\frac{\epsilon^{(k)}_{m+1}}{T}})\nonumber\\
& &-T\sum_{m=1}^{\infty}a_m*\left(\ln(1+e^{-\frac{\epsilon^{(k-1)}_m}{T}})+
\ln(1+e^{-\frac{\epsilon^{(k+1)}_m}{T}})\right),\label{TBAe1}\\
\epsilon^{(k)}_n&=&g_n^{(k)}+Ta_1*\ln(1+e^{\frac{\epsilon^{(k)}_{n-1}}{T}}+
Ta_2*\ln(1+e^{-\frac{\epsilon^{(k)}_n}{T}}))\nonumber\\
& &+T(a_0+a_2)\sum_{m\geq
  n}^{\infty}a_{m-n}*\ln(1+e^{-\frac{\epsilon^{(k)}_m}{T}})    \\
& &-T\sum_{m\geq n}^{\infty}a_{m-n+1}*\left(\ln(1+e^{-\frac{\epsilon^{(k-1)}_m}{T}})+
\ln(1+e^{-\frac{\epsilon^{(k+1)}_m}{T}})\right),
\,n\geq 2. \nonumber \label{TBAe2}
\end{eqnarray}
Here $\rho ^{(k)h}_n (\lambda)/\rho_n^{(k)}(\lambda) := \exp(\epsilon^{(k)}_n(\lambda)/T)$ 
with $k=1,\ldots,5$
and $\epsilon^{(0)}_n(\lambda)=\epsilon^{(6)}_n(\lambda)=0$ is
assumed.  The dressed energies $\epsilon^{(k)}_n$ play the role of
excitation energies measured from the Fermi level for each flavour.  The
driving terms in the antiferromagnetic rung coupling regime for a weak
magnetic field, $h<\frac{3J_{\perp} }{[g_t+3g_s+(g_s-g_t)/3]\mu_B}$, are given by
\begin{eqnarray} g_1^{(1)}&=&
-J_{\parallel}\frac{1}{v^2+\frac{1}{4}}+[g_s+\ffrac{1}{3}(g_s-g_t)]\mu_Bh, \nonumber \\
g_1^{(2)}&=&\ffrac{3}{2}J_{\perp}-[\ffrac{1}{2}g_t+\ffrac{3}{2}g_s+\ffrac{1}{6}(g_s-g_t)]\mu_Bh,
\nonumber\\
g_1^{(3)}&=&[\ffrac{1}{2}(g_t+g_s)+\ffrac{1}{6}(g_s-g_t)]\mu_Bh,\label{DT1} \\
g_1^{(4)}&=&[g_s-\ffrac{1}{3}(g_s-g_t)]\mu_Bh,\nonumber\\
g_1^{(5)}&=&[\ffrac{1}{2}(g_t+g_s)+\ffrac{1}{6}(g_s-g_t)]\mu_Bh. \nonumber 
\end{eqnarray}
The higher driving terms are given by $g_n^{(1)}=n[g_s+\frac{1}{3}(g_s-g_t)]\mu_Bh $ and  
$g_n^{(k)}=ng_1^{(k)}$ for $k >1$.  
Consequently, the free energy is given by
\begin{eqnarray} 
\frac{f(h,T)}{L}=
-\ffrac{1}{2}g_s\mu_Bh-\ffrac{1}{6}(g_s-g_t)\mu_Bh
-T\int_{-\infty}^{\infty}\sum_{n=1}^{\infty}a_n(\lambda)
\ln(1+e^{-\frac{\epsilon^{(1)}_n(\lambda)}{T}})d\lambda.
\end{eqnarray}
It is worth mentioning that the driving terms vary for different choices of the basis order.
The TBA equations (\ref{TBAe1}) and (\ref{TBAe2}) provide a clear physical picture of
the groundstate and make the thermodynamic properties, such as 
the free energy, magnetization and susceptibility accessible.

\section{Quantum phase diagram} \label{sec3}


In the low temperature limit, the states with positive dressed energy are
empty. The rapidities with negative dressed energy correspond to
occupied states.  The zeros of the dressed energies define the Fermi
energies. As usual, we decompose $\epsilon^{(a)}_n$ into positive
and negative parts,
$\epsilon^{(k)}_n=\epsilon^{(k)+}_n+\epsilon^{(k)-}_n$, with only the
negative dressed energies contributing to the groundstate energy. 
Analysis of equations (\ref{TBAe1}) and (\ref{TBAe2}) in the limit
$T\rightarrow 0$ reveals that the roots are all real for the groundstate, 
corresponding to $n=1$. 
All dressed energies $\epsilon^{(k)+}_n$ with $n\geq 2$ correspond to excitations. 
Under this circumstance, we see that all energy bands are completely filled in the absence 
of an external magnetic field and rung interactions.

In order to derive the groundstate properties, we first consider the
antiferromagnetic regime $J_{\perp}>0$, where the doublet component
$\psi_{\frac{1}{2}}^{(-)}$ is chosen as the reference state with a basis order
$(\psi_{\frac{1}{2}}^{(-)},\psi_{-\frac{1}{2}}^{(-)},\psi_{\frac{3}{2}},\psi_{\frac{1}{2}}^{(+)},
\psi_{-\frac{1}{2}}^{(+)},\psi_{-\frac{3}{2}})$.
The groundstate TBA equations then read
\begin{eqnarray}
\epsilon^{(1)}&=&g_1^{(1)}-a_2*\epsilon^{(1)-}+a_1*\epsilon^{(2)-},\nonumber\\
\epsilon^{(k)}&=&g^{(k)}_1-a_2*\epsilon^{(k)-}+a_1*\left[\epsilon^{(k-1)-}+
\epsilon^{(k+1)-}\right],\label{TBAg1}\\
&&k = 2,\ldots,5.\nonumber
\end{eqnarray}
It is clear that without a magnetic field the $su(6)$ multiplet levels split due to the rung coupling. 
If $J_{\perp}$ is very large (the limit of strong rung coupling), the whole quadruplet state 
$(\psi_{\frac{3}{2}},\psi_{\frac{1}{2}}^{(+)},\psi_{-\frac{1}{2}}^{(+)},\psi_{-\frac{3}{2}})$
is gapfull, i.e. $\epsilon^{(k)}>0$ for $k>2$.
This means that the quadruplet is not involved in the groundstate -- the
groundstate consists of doublet states with massless excitation.
Solving the TBA equation (\ref{TBAg1}), we find that the quadruplet excitation gap is given by
$\Delta_1=\frac{3}{2}J_{\perp}-2J_{\parallel}\ln 2$.
Thus if $J_{\perp}$ becomes larger than the critical value
$J_c^+=\frac{4}{3}J_{\parallel}\ln 2$, there is a quantum phase transition
from the Luttinger liquid $su(4)\oplus su(2)$ phase into the $su(2)$ phase.
However, in the presence of a magnetic field this critical point is not stable. 
In this case the dressed energy levels become completely split by the magnetic field $h$. 
If the rung coupling $J_{\perp}$ is large enough so that the driving term $g_1^{(2)}$ in
(\ref{DT1}) remains positive, the quadruplet state could be gapfull and the
groundstate would still be the doublet.
However, as the magnetic field increases, the doublet component $\psi_{-\frac{1}{2}}^{(-)}$ gradually
shifts out of the groundstate as  the Fermi surface of the dressed energy $\epsilon^{(1)}$ lifts.
Subsequently, if  the magnetic field $h$ is larger than the critical point $H_{c1}$, the reference state
becomes a true physical state such that the strong  rung coupling forms a rung trimerized groundstate
(see Fig. 2). 
In other words, the doublet component  $\psi_{\frac{1}{2}}^{(-)}$ forms a ferromagnetic groundstate.
The critical field $H_{c1}$ is given by
\begin{equation}
H_{c1}=
4J_{\parallel}/[g_s+\ffrac{1}{3}(g_s-g_t)]\mu_B.\label{Hc1}
\end{equation}

\begin{figure}[t]
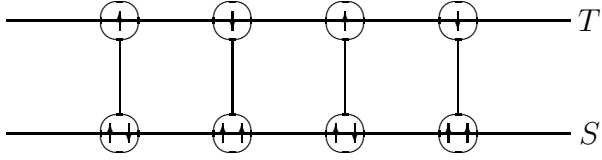

\begin{center}
\LaddF \\
\end{center}
\caption{The doublet polarized state forms a rung trimerized  groundstate. }
\end{figure}

The critical point $H_{c1}$ indicates a quantum phase
transition from a gapless magnetic phase into a ferromagnetic phase
with gap {$\Delta_2=(\frac{1}{2}g_s+\frac{1}{6}(g_s-g_t)]\mu_B(h-H_{c1})$}.
It is worth noting that in this ferromagnetic phase a magnetization plateau
$M^z=\frac{1}{2} g_s\mu_B+\frac{1}{6}(g_s-g_t)\mu_B$ opens.
The necessary condition for this  plateau to exist is
\begin{eqnarray}
J_{\perp} \geq J_c^{+F}=\frac{8J_{\parallel}}{3}\frac{[\frac{1}{2}g_t+\frac{3}{2}g_s+
\frac{1}{6}(g_s-g_t)]}{[g_s+\frac{1}{3}(g_s-g_t)]}.\label{Jc}
\end{eqnarray}
In the critical phase  $h< H_{c1}$,
 the TBA equations with  a very large or a very
small Fermi boundary can be solved  analytically. If the magnetic
field is very small, i.e. $ h\!\ll\! 1$, the Fermi boundary of the dressed
energy $\epsilon ^{(1)}$ is very large.
The energy potential satisfies the Wiener-Hopf type equation 
\begin{equation}
\epsilon^{(1)}(\lambda)=-J_{\parallel}\frac{\pi }{\cosh \pi
\lambda}+\ffrac{1}{2}[g_s+\ffrac{1}{6}(g_s-g_t)]\mu_Bh+\int_{-\infty,B}^{-B,\infty}
G(\lambda -k)*\epsilon^{(1)}(k)dk.
\end{equation}
Here the function $G(\lambda)$ is defined via
\begin{eqnarray}
G(\lambda)
=\frac{1}{2\pi}\int_{-\infty}^{\infty}\frac{e^{-|\omega|/2}}{2\cosh\omega/2}e^{-{\mathrm
    i}\lambda \omega}d\omega.\nonumber
\end{eqnarray}
Using the standard Wiener-Hopf technique, we find that  the Fermi boundary
satisfies the relation
$e^{-B\pi}=[g_s+\frac{1}{3}(g_s-g_t)]\mu_Bh a_-(0)/4J_{\parallel}\pi
  a_+(\mathrm{i}\pi)$, where
the decomposition functions are given by
\begin{equation}
a_+(\omega)=a_-(-\omega)=\sqrt{2}\pi \left(\frac{\eta -i\omega }{2\pi
    e}\right)^{-\frac{\mathrm{i}\omega }{2\pi}}\!\!\!/\Gamma
    (\frac{1}{2}-\frac{\mathrm{i}\omega }{2\pi}).
\end{equation}
The Fermi boundary decreases monotonically with increasing magnetic field. 
Correspondingly, the free energy is given by
\begin{equation}
\frac{F(0,h)}{L}\approx -J_{\parallel}[\Psi(1)-\Psi(\ffrac{1}{2})]-\frac{1}{8\pi ^2}
[g_s+\ffrac{1}{3}(g_s-g_t)]^2\mu^2_Bh^2,
\end{equation}
which suggests a susceptibility of 
$\chi \approx \frac{1}{4\pi ^2}[g_s+\frac{1}{3}(g_s-g_t)]^2\mu _B^2$, 
indicating an $su(2)$ critical phase. In the above $\Gamma (z)$ and $\Psi(a)$ are the gamma and 
diagamma functions, respectively.

On the other hand, if the magnetic field $h$
tends to the critical point $H_{c1}$, the Fermi boundary $Q$ of the
dressed energy $\epsilon ^{(1)}$ is very small, say $Q\!\ll \! 1$ for $H_{c1}\!-\!h\!\ll \! 1$.
Under this circumstance, the free energy is given by
\begin{equation}
\frac{F(0,h)}{L}\approx -\ffrac{1}{2}g_s\mu_B-\ffrac{1}{6}(g_s-g_t)\mu_Bh-\frac{4Q}
{\pi}[g_s+\ffrac{1}{3}(g_s-g_t)]\mu_B(H_{c1}-h),
\end{equation}
where $Q\approx \sqrt{(H_{c1}-h)/4H_{c1}}$.
Thus the susceptibility
\begin{equation}
\chi \approx \frac{(4g_s-g_t) \mu_B}{\pi \sqrt{4H_{c1}}}(H_{c1}-h)^{-\frac{1}{2}}
\end{equation}
indicates the singular behavior of the transition from the gapless phase into the gapped phase.
In addition, from the Bethe ansatz equations and the relation
\begin{equation}
M^z\approx \ffrac{1}{2}g_s\mu_B+\ffrac{1}{6}(g_s-g_t)\mu_B-[g_s+\ffrac{1}{3}(g_s-g_t)]
\mu_B\int_{-Q}^{Q}\rho^{(1)}_1(\lambda)d\lambda,
\end{equation}
the magnetization per site  $M^z$ in the vicinity of $H_{c1}$ follows as
\begin{equation}
M^z=\ffrac{1}{2}g_s\mu_B+\ffrac{1}{6}(g_s-g_t)\mu_B-\frac{4Q}{\pi}[g_s+\ffrac{1}{3}(g_s-g_t)] 
\mu_B(1-\frac{2Q}{\pi}).\label{MZ1}
\end{equation}
Apparently, as $h\!\!\to\!\! H_{c1}$ the magnetization $M^z$ tends to the
plateau value $\frac{1}{2}g_s \mu_B+\frac{1}{6}(g_s-g_t)\mu_B$. We shall discuss the magnetization
plateaux as well as the quantum phase transitions in the next section.

For the ferromagnetic regime $J_{\perp}<0$, the rung quadruplet component $\psi_{\frac{3}{2}}$ 
is chosen as the reference state with
$(\psi_{\frac{3}{2}},\psi_{\frac{1}{2}}^{(+)},\psi_{-\frac{1}{2}}^{(+)},\psi_{-\frac{3}{2}},\psi_{\frac{1}{2}}^{(-)},
\psi_{-\frac{1}{2}}^{(-)})$
as the order of the basis.
Thus the driving terms are given by
\begin{eqnarray}
g_1^{(1)}&=&-J_{\parallel}\frac{1}{v^2+\frac{1}{4}}+[\ffrac{1}{2}(g_t+g_s)+\ffrac{1}{6}(g_s-g_t)]\mu_Bh,\nonumber\\
g_1^{(2)}&=&[g_s-\ffrac{1}{3}(g_s-g_t)]\mu_Bh, \nonumber\\
g_1^{(3)}&=&[\ffrac{1}{2}(g_t+g_s)+\ffrac{1}{6}(g_s-g_t)]\mu_Bh, \\
g_1^{(4)}&=&-\ffrac{3}{2}J_{\perp}-[\ffrac{1}{2}g_t+\ffrac{3}{2}g_s+\ffrac{1}{6}(g_s-g_t)]\mu_Bh,\nonumber\\
g_1^{(5)}&=&[g_s+\ffrac{1}{3}(g_s-g_t)]\mu_Bh.\nonumber
\end{eqnarray}

In the absence of a magnetic field, the quadruplet and doublet states are
degenerate. If the rung coupling becomes negative enough
the doublet state is completely gapfull, and the Fermi
boundaries of the quadruplet states are infinity.
Using Fourier transforms, we find that the doublet does not exist in the groundstate
for $J_{\perp} < J_{c}^-=-\frac{2}{3} J_{\parallel}$.
Again this critical point is not stable if the magnetic field is applied. 
We can see that the Fermi surfaces of the dressed energies $\epsilon^{(k)},\,k=1,2,3 $ lift,
while the Fermi surfaces  of the doublet sink.  
If the magnetic field is strong enough, i.e. if
\begin{equation}
h\geq h_c^{-}=4J_{\parallel}/[\ffrac{1}{2}(g_t+g_s)+\ffrac{1}{6}(g_s-g_t)]\mu_B,
\end{equation}
the reference state $\psi _{\frac{3}{2}}$ becomes a true  physical state. 
Thus  the groundstate is a fully-polarized (trimer-like) ferromagnetic state (see Fig. 3).
Note that the rung coupling $J_{\perp}$ must be less than a critical value  $J_{\perp}^{-F} $ given by
\begin{equation}
J_{\perp}^{-F}=-\frac{8J_{\parallel}}{3}\frac{[\frac{1}{2}g_t+\frac{3}{2}g_s+\frac{1}{6}(g_s-g_t)]}{[\frac{1}{2}(g_t+g_s)+\frac{1}{6}(g_s-g_t)]}.
\end{equation}

\begin{figure}[t]
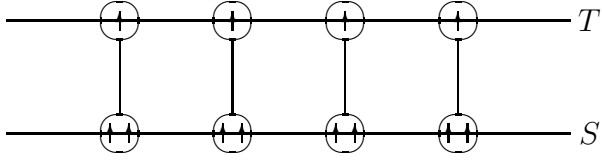

\begin{center}
\Ladderg2 \\
\end{center}
\caption{The quadruplet fully-polarized state forms a trimerized ferromagnetic groundstate. }
\end{figure}

\section{Magnetization plateaux}
\label{sec4}

Magnetization plateaux are one of the most interesting phenomena in the ladder compounds. 
For example, fractional  magnetization plateaux have been found in Shastry-Sutherland systems \cite{FMP}.
Theoretical studies and numerical results suggest that magnetization plateaux exist in the mixed
spin-$(\frac{1}{2},1)$ chains \cite{mix1,mix2,mix3} and the mixed ladder \cite{mlad1}. 
From the analysis of the critical points of the solvable model in the last section, we found
that gapped or gapless states appear in turn as the external magnetic field increases. 
For very strong rung coupling, i.e. $J_{\perp}\!\gg\! J_{c}^{+F}$, the
two-component massless quantum magnetic phase lies in the regime $h\!<\!H_{c1}$.
The ferromagnetic phase appears for a magnetic field $h\!>\!H_{c1}$ and the component
$\psi^{(-)}_{\frac{1}{2}}$ becomes a physical ferromagnetic groundstate. 
On the other hand, the magnetic field can bring the state $\psi_{\frac{3}{2}}$ 
close  to the  groundstate.
Eventually it  becomes involved in the groundstate when the magnetic field is strong enough. 
From the expression (\ref{energy}), we see that if
$h>3J_{\perp}/(g_t+3g_s+\frac{1}{3}(g_s-g_t))\mu_B$, the state $\psi_{\frac{3}{2}}$  becomes an
energetically lower lying state than $\psi_{-\frac{1}{2}}^{(-)}$.  
In this case it is convenient to reorder the basis as 
$(\psi_{\frac{1}{2}}^{(-)},\psi_{\frac{3}{2}},\psi_{-\frac{1}{2}}^{(-)},\psi_{\frac{1}{2}}^{(+)},
\psi_{-\frac{1}{2}}^{(+)},\psi_{-\frac{3}{2}})$.
Here the TBA  driving terms are given by
\begin{eqnarray}
g_1^{(1)}&=&-J_{\parallel}\frac{1}{v^2+\frac{1}{4}}+\ffrac{3}{2} J_{\perp}-
[\ffrac{1}{2}(g_t+g_s)-\ffrac{1}{6}(g_s-g_t)]\mu_Bh, \nonumber\\
g_1^{(2)}&=&-\ffrac{3}{2}J_{\perp}+[\ffrac{1}{2}g_t+\ffrac{3}{2}g_s+
\ffrac{1}{6}(g_s-g_t)]\mu_Bh,\nonumber\\
g_1^{(3)}&=&\ffrac{3}{2}J_{\perp}-g_s\mu_Bh, \\
g_1^{(4)}&=&[g_s-\ffrac{1}{3}(g_s-g_t)]\mu_Bh,\nonumber\\
g_1^{(5)}&=&[\ffrac{1}{2}(g_t+g_s)+\ffrac{1}{6}(g_s-g_t)]\mu_Bh. \nonumber 
\end{eqnarray}

Analysing the TBA with these driving terms,  we find  that the ferromagnetic groundstate can be 
 maintained only in the regime $H_{c1}\leq h<H_{c2}$, where the critical point $H_{c2}$ is given by
\begin{equation}
H_{c2}=\frac{3J_{\perp}-8J_{\parallel}}{[(g_t+g_s)-\frac{1}{3}(g_s-g_t)]\mu_B}.\label{Hc2}
\end{equation}
Beyond the critical field $H_{c2}$, the plateau $M^z=\frac{1}{2}
g_s\mu_B+\frac{1}{6}(g_s-g_t)\mu_B$  vanishes. In the vicinity of $H_{c2}$, i.e. $h-H_{c2}\!\ll\!1$, the Fermi
boundary is very small. After a similar calculation the Fermi point is found to be
\begin{equation}
Q\approx \sqrt{\frac{[\frac{1}{2}(g_t+g_s)-\frac{1}{6}(g_s-g_t)]\mu_B(h-H_{c2})}{16J_{\parallel}}}.
\end{equation}
The susceptibility is given by
\begin{equation}
\chi \approx \frac{3[\frac{1}{2}(g_s+g_t)\mu_B-\frac{1}{6}(g_s-g_t)\mu_B]^{\frac{3}{2}}}{4\pi \sqrt{J_{\parallel}(h-H_{c2})}},
\end{equation}
 which indicates the singular behavior of the transition of the gapped phase into the gapless phase.
The magnetization shows the square root field dependent behaviour
\begin{equation}
M^z=[\ffrac{1}{2}g_s+\ffrac{1}{6}(g_s-g_t)]\mu_B+\frac{4Q}{\pi}[\ffrac{1}{2}(g_t+g_s)-\ffrac{1}{6}(g_s-g_t)]\mu_B(1-\frac{2Q}{\pi}). \label{MZ2}
\end{equation}

If the magnetic field is further increased
to  $h>3J_{\perp}/[(g_t+g_s)-\frac{1}{3}(g_s-g_t)]\mu_B$,  the state $\psi _{\frac{3}{2}}$ becomes the lowest lying state.
Thus in this regime, it is reasonable  to choose the basis order as
 $(\psi_{\frac{3}{2}},\psi_{\frac{1}{2}}^{(-)},\psi_{\frac{1}{2}}^{(+)},\psi_{-\frac{1}{2}}^{(-)},\psi_{-\frac{1}{2}}^{(+)},\psi_{-\frac{3}{2}})$. 
Subsequently, the driving terms are given by
\begin{eqnarray}
g_1^{(1)}&=&-J_{\parallel}\frac{1}{v^2+\frac{1}{4}}-\ffrac{3}{2} J_{\perp}+[\ffrac{1}{2}(g_t+g_s)-
\ffrac{1}{6}(g_s-g_t)]\mu_Bh, \nonumber\\
g_1^{(2)}&=&\ffrac{3}{2} J_{\perp} +\ffrac{1}{3}(g_s-g_t)]\mu_Bh,\nonumber\\
 g_1^{(3)}&=&-\ffrac{3}{2} J_{\perp} +g_s\mu_Bh, \label{DT4} \\
g_1^{(4)}&=&\ffrac{3}{2} J_{\perp} -\ffrac{1}{3}(g_s-g_t)\mu_Bh\nonumber,\\
g_1^{(5)}&=&[\ffrac{1}{2}(g_t+g_s)+\ffrac{1}{6}(g_s-g_t)]\mu_Bh. \nonumber  
\end{eqnarray}
As the magnetic field increases beyond  the critical point  $H_{c2}$, the groundstate
becomes a mixture of the doublet and quadruplet states.
Strictly speaking, the doublet component $\psi_{\frac{1}{2}}^{(-)}$ and the quadruplet component
$\psi _{\frac{3}{2}}$ compete for the groundstate. 
Other components of the multiplets are gapfull by virtue of both the rung coupling  and  magnetic field.
As  the magnetic field becomes stronger, the probability of the quadruplet component 
$\psi _{\frac{3}{2}}$ becomes higher. The inflection point at  
$h_{{\rm IP}}=3J_{\perp}/[(g_t+g_s)-\frac{1}{3}(g_s-g_t)]\mu_B$ indicates an equal  
probability between the components $\psi _{\frac{3}{2}}$ and $\psi _{\frac{1}{2}}^{(-)}$, 
which can be seen clearly from the magnetization curve in Fig. 4.
Using the TBA equations with the driving term (\ref{DT4}), we find
that for a sufficiently large magnetic field $h\geq H_{c3}$ the groundstate 
becomes fully-polarized  with a full saturation magnetization plateau at 
$M^z=(\frac{1}{2}g_t+g_s)\mu_B$. 
The critical point is given by
\begin{eqnarray}
H_{c3}=\frac{3J_{\perp}+8J_{\parallel}}{[(g_t+g_s)-\frac{1}{3}(g_s-g_t)]\mu_B}. \label{Hc3}
\end{eqnarray}
Analogously, we find the singular behavior in the vicinity of the the critical point $H_{c3}$.
The susceptibility is given by
\begin{equation}
\chi \approx \frac{3[\frac{1}{2}(g_s+g_t)\mu_B-\frac{1}{6}(g_s-g_t)\mu_B]^{\frac{3}{2}}}{4\pi \sqrt{J_{\parallel}(H_{c3}-h)}},
\end{equation}
which indicates the nature of the singular behavior of the transition between the gapless and gapped phases.
The magnetization also exhibits the square root field dependent behaviour
\begin{equation}
M^z=[\ffrac{1}{2}g_t+g_s]\mu_B-\frac{4Q}{\pi}[\ffrac{1}{2}(g_t+g_s)-\ffrac{1}{6}(g_s-g_t)]
\mu_B(1-\frac{2Q}{\pi}),\label{MZ3}
\end{equation}
where the Fermi point $Q$ is 
\begin{equation}
Q\approx
\sqrt{\frac{[\frac{1}{2}(g_t+g_s)-\frac{1}{6}(g_s-g_t)]\mu_B(H_{c3}-h)}{16J_{\parallel}}}.
\end{equation}
The magnetization increases almost linearly between the critical fields $H_{c2}$ and $H_{c3}$.

\begin{figure}
\centerline{
\epsfig{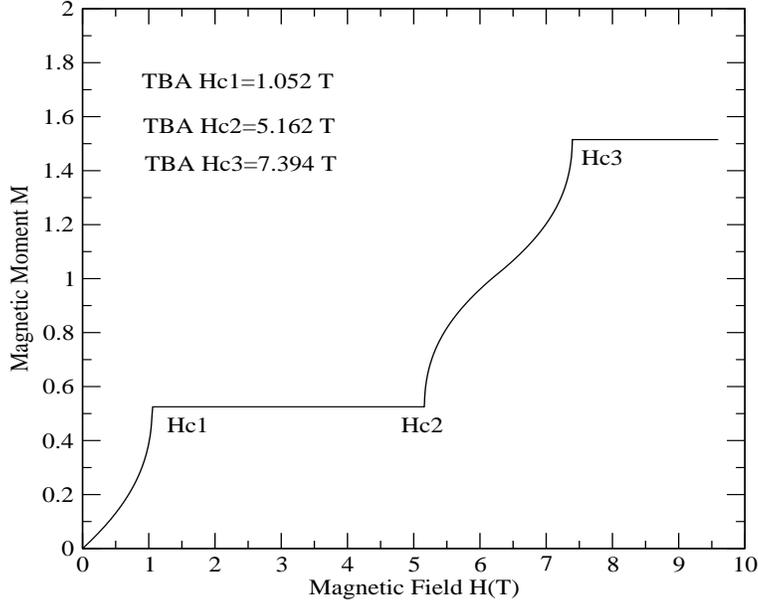}}
\caption{The magnetization versus magnetic field $h$ in the strong antiferromagnetic 
rung coupling regime. 
The magnetic moment is normalized from the magnetization via $M^z/\frac{1}{2}(g_s+g_t)\mu_B$. 
The coupling constants used are $J_{\perp}=6.0K$, $J_{\parallel}=0.4K$ and we take  
$g$-factor values  $g_s=2.22,\,g_t=2.09$  with $\mu_B=0.672 K/T$.
From the TBA we predict that the one third saturation magnetization plateau opens only if 
$J_{\perp} \geq J_c^{+F}$ as given in (\ref{Jc}).
The indicated critical fields $H_{c1}\approx 1.052 $ T, $H_{c2}\approx 5.162 $ T and 
$H_{c3}\approx 7.394 $ T predicted by the TBA coincide with the numerically estimated values.
The inflection point at  $h=h_{{\rm IP}}\approx 6.278$ T and  $M \approx 1$  indicates a 
point of equal probability for the states $\psi_{\frac{3}{2}}$ and $\psi_{\frac{1}{2}}^{(-)}$. }
\label{Fig1}
\end{figure}

\begin{figure}
\centerline{
\epsfig{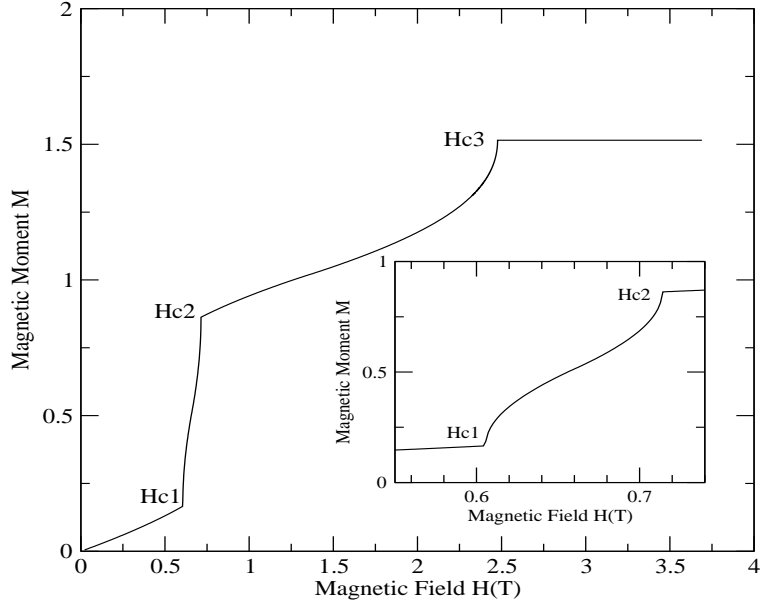}}
\caption{The magnetization versus magnetic field $h$ in the weak antiferromagnetic rung coupling regime.
The numerical values are the same as for the previous figure, but now using the smaller rung
coupling constant $J_{\perp}=1.3K<J_c^{+ F}\approx 2.07K$. In this case the fractional
magnetization plateau vanishes.
The TBA critical fields coincide again with the numerically estimated values.
The inset shows an enlargement of the magnetization between $H_{c1}$ and $H_{c2}$.}
\label{Fig2}
\end{figure}

We have obtained the whole  magnetization curve by numerically solving the TBA equations 
in the different phases (see Fig.\ref{Fig1}). 
A third and full saturation magnetization plateaux is observed.
On the other hand, as mentioned above, the first magnetization
plateau at $M^z=\frac{1}{2}g_s\mu_B+\frac{1}{6}(g_s-g_t)\mu_B$ depends mainly on the rung coupling.
If $J_{\perp}<J_c^{+F}$ this plateau disappears. 
$J_c^{+F}$ is a necessary condition for the existence of the one third saturation 
magnetization  plateau. 
This conclusion is reasonable because the leg part of the Hamiltonian (\ref{Ham})
is stronger than the rung part, due to the multispin interaction terms in eq.(\ref{intra}). 
Therefore, the rung coupling must be large enough in order to dominate the low temperature groundstate.
If the rung coupling fulfills $J_c^+\!<\!J_{\perp}\! <\!J_c^{+F}$ the 
field-induced fractional plateau vanishes.
In Fig. \ref{Fig2} we show the full numerical magnetization curve for 
weak rung coupling $J_{\perp}=1.3 $K.
We clearly see that the fractional magnetization plateau is closed. 
From the TBA analysis the first critical 
point $H_{c1}$ lies in the interval
\begin{eqnarray}
\frac{3(J_{\perp}-J_c^+)}{[g_t+3g_s+(g_s-g_t)/3]\mu_B}  
<H_{c1}\!<\!\frac{3J_{\perp}}{[g_t+3g_s+(g_s-g_t)/3]\mu_B},\nonumber
\end{eqnarray}
which for the given  parameter set (see figure captions) evaluates to $0.47\!<\!H_{c1}\!<\!0.66$ T.
Indeed, from the numerical results we find $H_{c1} \approx 0.605 $ T.
This implies that for $H< 0.605$T the groundstate is the doublet spin liquid phase. 
A magnetic field beyond this point allows the quadruplet component $\psi _{\frac{3}{2}}$ 
to be involved in the groundstate. Therefore the critical point $H_{c1}$ indicates a
quantum phase transition from a two-state phase into a three-state phase. 
Hence for $h\!>\!0.605 $ T, two Fermi seas, $\epsilon ^{(1)}$ and $\epsilon ^{(2)}$, 
lie in the groundstate.  We can see that the probability of the component 
$\psi _{\frac{3}{2}}$ to be in the groundstate increases  as the magnetic field increases.
Meanwhile the doublet state $\psi^-_{-\frac{1}{2}}$ is quickly driven out of the 
groundstate at the critical point $H_{c2}$, where the phase transition from the 
three-state into the two-state phase transition occurs.
From Eq. (\ref{energy}), we find  the middle point between $H_{c1}$ and $H_{c2}$ is 
approximately at $0.66$ T,
which is consistent with the numerical curve in Fig. \ref{Fig2}. 
Thus from the theory  we predict the critical field to be  
$H_{c2}=6J_{\perp}/[(g_t+g_s)-\frac{1}{3}(g_s-g_t)]\mu_B-H_{c1}
\approx 0.715$ T.  This again is in agreement with the numerical value of $0.713 $ T. 
In the region $H_{c2}\!<\!h\!<\!H_{c3}$ the two components 
$\psi^-_{\frac{1}{2}}$ and $\psi _{\frac{3}{2}}$
compete to be in the groundstate. If the magnetic field is strong enough, so that $H>H_{c3}$
where $H_{c3}$ is given by (\ref{Hc3}), the reference state 
$\psi _{\frac{3}{2}}$ becomes a true physical groundstate.

\section{Conclusion} \label{sec5}

We have investigated the phase diagram of the exactly solved mixed spin-$(\frac{1}{2},1)$ ladder
model in both the absence and presence of an external magnetic field using the TBA.
It has been shown that in the strong antiferromagnetic rung coupling regime there exists
a third and full saturation magnetization plateaux.  
A Luttinger liquid magnetic phase exists in the regime $h<H_{c1}$, which corresponds to the
doublet $su(2)$ phase.
The magnetic  groundstate consisting of two components  $\psi^-_{\frac{1}{2}}$ and $\psi _{\frac{3}{2}}$
lies in the regime  $H_{c2}\!<\!h\!<\!H_{c3}$.
The ferromagnetic ground state with a third saturation magnetization plateau appears in the
regime $H_{c1}\!<\!h\!<\!H_{c2}$.
The full saturation magnetization plateau opens at  $h>\!H_{c3}$.
The gapped or gapless states appear in turn as the magnetic field is increased.
The weak rung coupling regime exhibits three different phase transitions,
which involve two- and three-state quantum phase transitions.
The fractional magnetization plateau vanishes.
The model does not exihibit a third saturation magnetization plateau in the
strong ferromagnetic rung coupling regime.
We have also investigated the singular behaviour in the vicinity of the critical points 
via the solutions of the TBA equations.
As the contributions from the leg interaction to the groundstate energy are very small in the
strong rung coupling regime we believe that  the solvable model (\ref{Ham}) is well suited to describe
the physics of real mixed spin-($\frac{1}{2},1$) ladder compounds with a relatively large rung 
coupling constant.
However, such mixed spin ladder compounds are yet to be found.

One compound that we are aware of is the organic ferrimagnet PNNBNO \cite{mixladd},
which has been recognized as a ladder compound with alternating spin-$\frac{1}{2}$ and
spin-$1$ units, i.e. as two coupled alternating mixed spin chains.
The strong interchain (rung) coupling suggests that PNNBNO can be effectively 
identified as a mixed spin-($\frac{1}{2},1$) ladder model via the Hamiltonian 
(\ref{Ham}) in the high temperature limit ($T \geq 50$ K). 
In this case the dominant rung interaction forms an effective one-dimensional 
spin-$\frac{3}{2}$ antiferromagnetic chain. 
Of course for low temperatures the ferrimagnetic correlations between the 
spin-$1$ and spin-$\frac{1}{2}$ units should appear.
Nevertheless we believe that the mixed spin-($\frac{1}{2},1$) ladder model 
discussed here can at least describe the high temperature properties of  
PNNBNO \cite{mixladd}.

\vskip 1cm
{\bf Acknowledgments}.
This work has been supported by the Australian Research Council. ZJY thanks FAPERGS
for financial support.
We thank A. Foerster and  H.-Q. Zhou for helpful discussions. XWG also thanks the 
Centre for Mathematical Physics at the University of Queensland for kind hospitality.

\end{document}